\documentclass[conference]{IEEEtran}
\usepackage{cite}
\usepackage{amsmath,amssymb,amsfonts}
\usepackage{algorithmic}
\usepackage{graphicx}
\usepackage{textcomp}
\usepackage{xcolor}
\def\BibTeX{{\rm B\kern-.05em{\sc i\kern-.025em b}\kern-.08em
    T\kern-.1667em\lower.7ex\hbox{E}\kern-.125emX}}

\usepackage{todonotes}
\usepackage{xurl}
\usepackage{listings}
\usepackage{booktabs}
\usepackage{hyperref}
\usepackage{enumitem}

\definecolor{codegray}{rgb}{0.95,0.95,0.95} % light gray background
\definecolor{codeblue}{rgb}{0,0,0.6}
\definecolor{codegreen}{rgb}{0,0.5,0}
\definecolor{codered}{rgb}{0.6,0,0}

\begin{document}

\title{Dynamic Resource Management in Production HPC Clusters}

%\author{\IEEEauthorblockN{Anonymous Author(s)}}

\author{\IEEEauthorblockN{1\textsuperscript{st} Petter Sandås}
\IEEEauthorblockA{\textit{Barcelona Supercomputing Center} \\
\textit{(BSC)}\\
Barcelona, Spain \\
0009-0009-2028-7719}
\and
\IEEEauthorblockN{2\textsuperscript{nd} Sergio Iserte}
\IEEEauthorblockA{\textit{Barcelona Supercomputing Center} \\
\textit{(BSC)}\\
Barcelona, Spain \\
0000-0003-3654-7924}
\and
\IEEEauthorblockN{3\textsuperscript{rd} Guillaume Houzeaux}
\IEEEauthorblockA{\textit{Barcelona Supercomputing Center} \\
\textit{(BSC)}\\
Barcelona, Spain \\
0000-0003-3654-7924}
\and
\IEEEauthorblockN{4\textsuperscript{th} Antonio J. Peña}
\IEEEauthorblockA{\textit{Barcelona Supercomputing Center} \\
\textit{(BSC)}\\
Barcelona, Spain \\
0000-0002-3575-4617}
}

\iffalse
\author{\IEEEauthorblockN{1\textsuperscript{st} Given Name Surname}
\IEEEauthorblockA{\textit{dept. name of organization (of Aff.)} \\
\textit{name of organization (of Aff.)}\\
City, Country \\
email address or ORCID}
\and
\IEEEauthorblockN{2\textsuperscript{nd} Given Name Surname}
\IEEEauthorblockA{\textit{dept. name of organization (of Aff.)} \\
\textit{name of organization (of Aff.)}\\
City, Country \\
email address or ORCID}
\and
\IEEEauthorblockN{3\textsuperscript{rd} Given Name Surname}
\IEEEauthorblockA{\textit{dept. name of organization (of Aff.)} \\
\textit{name of organization (of Aff.)}\\
City, Country \\
email address or ORCID}
\and
\IEEEauthorblockN{4\textsuperscript{th} Given Name Surname}
\IEEEauthorblockA{\textit{dept. name of organization (of Aff.)} \\
\textit{name of organization (of Aff.)}\\
City, Country \\
email address or ORCID}
\and
\IEEEauthorblockN{5\textsuperscript{th} Given Name Surname}
\IEEEauthorblockA{\textit{dept. name of organization (of Aff.)} \\
\textit{name of organization (of Aff.)}\\
City, Country \\
email address or ORCID}
\and
\IEEEauthorblockN{6\textsuperscript{th} Given Name Surname}
\IEEEauthorblockA{\textit{dept. name of organization (of Aff.)} \\
\textit{name of organization (of Aff.)}\\
City, Country \\
email address or ORCID}
}
\fi

\maketitle

\begin{abstract}
Many large-scale scientific applications exhibit time-varying behavior, yet production HPC clusters still rely on rigid, fixed-size allocations, and most dynamic techniques remain confined to laboratory prototypes. This work presents a practical MPI malleability methodology that integrates with state-of-the-art high-performance computing (HPC) software stacks and operational practices. The methodology is implemented in the Dynamic Management of Resources (DMR) framework and is designed to ease adoption by existing applications without requiring intrusive code changes or scheduler modifications. We evaluate our approach by integrating the DMR API into two large-scale scientific applications and deploying them on three TOP500 supercomputers under realistic production configurations. Our non-invasive malleability solution achieves performance comparable to static baselines in controlled environments while substantially reducing node-hour consumption for identical workloads. These results show that malleability can be effectively exploited on production systems using vanilla resource managers, lowering the barrier to adoption of dynamic resource management in HPC.
\end{abstract}


\section{Introduction}\label{sec:intro}
Many large-scale scientific applications exhibit behaviors that change significantly during execution, when, for example, individual physics components in a coupled workflow scale differently; surrogate models are offloaded to accelerators; or iterative solvers require a variable number of iterations per time step. Despite this intrinsic variability, most production HPC systems still rely on rigid, fixed-size allocations, and dynamic techniques are typically exercised only in controlled testbeds rather than day-to-day production deployments.

Modern high-performance computing (HPC) platforms can deploy hundreds of thousands of processing units, serving domains such as climate modeling, nuclear physics, genomics, and large-scale machine learning. This scale amplifies the need for dynamic and efficient use of resources, as mismatches between static allocations and time-varying workloads may lead to significant waste of compute capacity~\cite{balaprakash_exascale_2013}. Resource management systems (RMSs) allocate and schedule resources across thousands of concurrent jobs and are therefore central to enabling any form of dynamic workload management.

Dynamic resource management techniques enable jobs to acquire, release, or reconfigure resources at runtime, improving utilization and responsiveness in production environments~\cite{teich_invasive_2010}. Among these techniques, process malleability based on the Message Passing Interface (MPI) has attracted particular attention: MPI malleability allows running applications to reshape their process layout and redistribute data without user intervention~\cite{feitelson_packing_1996}. This remains an active area of research, with recent surveys and frameworks highlighting a growing ecosystem of libraries that aim to make MPI malleability more practical and widely usable~\cite{aliaga_survey_2022, tarraf_malleability_2024}.

In modern HPC environments, Slurm~\cite{yoo_slurm_2003} has emerged as a dominant RMS and workload scheduler, powering a substantial fraction of TOP500\footnote{\url{https://www.top500.org}} supercomputers, including several flagship exascale systems. It is commonly estimated that around 60\%–65\% of the systems in the TOP500 list run Slurm or closely related variants as their primary RMS\footnote{\url{https://bit.ly/hpcc-slurm}}, although this figure cannot be independently verified because many systems do not publicly disclose detailed configurations\footnote{\url{https://bit.ly/slurm-hpc-scheduler-for-workloads}}.

Despite its maturity and scalability, Slurm’s resource allocation paradigm remains largely static: jobs are typically bound to a fixed set of resources for their entire runtime, offering limited adaptability to time-varying workload characteristics and inter-job resource contention. Slurm does provide mechanisms\footnote{\url{https://slurm.schedmd.com/faq.html\#job_size}} to shrink a running job’s allocation via runtime updates to the node set or node count (e.g., using \texttt{scontrol update JobId=\# NumNodes=\#} or \texttt{NodeList=...}). However, these mechanisms operate only at the scheduler level; it is the application’s responsibility to reconfigure itself after a shrink, and most production MPI applications are not designed to handle such changes, often leading to failures or undefined behavior when resources are released during execution\footnote{\url{https://bit.ly/adaptive-multi-tier-exascale}}. As a consequence, these limitations hinder practical use of dynamic features and contribute to inefficient utilization of valuable compute capacity.

Some custom Slurm forks and experimental extensions support more advanced dynamic allocations and malleability, but cluster administrators are understandably hesitant to deploy non-vanilla releases due to concerns about long-term support, compatibility with upstream Slurm, and operational risk in production environments. As a result, many innovations in dynamic scheduling and malleable resource management remain confined to laboratory prototypes, with limited adoption in large-scale production systems.

To address these barriers, we propose a non-invasive mechanism for dynamic resource management in HPC clusters. 
%Although we focus on Slurm, the methodology is also applicable to other traditional RMSs such as PBS\footnote{\url{https://altair.com.es/pbs-professional/}} and exascale-oriented systems such as Flux\footnote{\url{https://flux-framework.org}}, since it does not rely on scheduler-specific invasive changes. 
Our approach enables real-time, user-transparent reallocation of compute resources, adapting to runtime conditions without altering the underlying RMS installation or requiring substantial restructuring of application code or job submission processes. By maintaining full compatibility with vanilla RMS deployments, the method offers a practical path to dynamic optimization in production HPC systems, particularly in exascale environments characterized by high resource heterogeneity and strong variability in job behavior.

The mechanism presented in this paper is implemented on top of the Dynamic Management of Resources (DMR, pronounced ``\emph{dimmer}'') malleability framework~\cite{iserte_agut_high-throughput_2018}. Our design enables users to deploy malleable codes in controlled environments with a malleability-capable Slurm configuration on dedicated clusters and, in addition, to run the same codes transparently on production systems, where Slurm is not configured to support malleability and users lack administrative privileges.

This paper makes the following contributions:
\begin{itemize}
    \item \textbf{Non-invasive model for MPI malleability over static RMSs}: we describe an orchestration methodology that decouples MPI malleability from RMS-level job resizing and coordinates expansion and shrink operations across multiple independent allocations using only user-level interactions with an unmodified scheduler.
    \item \textbf{Production-oriented implementation in DMRv2}: we extend DMR with a high-level API that supports both checkpoint/restart (C/R) and in-memory reconfiguration, enabling runtime resource reallocation for MPI applications on vanilla Slurm deployments without custom forks or experimental scheduler patches.
    \item \textbf{Evaluation on large production systems}: we integrate the API into two large-scale scientific applications and deploy them on three TOP500 supercomputers, demonstrating comparable performance to static baselines and substantial reductions in node-hour consumption under realistic production configurations.
\end{itemize}

The rest of the paper is structured as follows.
Section~\ref{sec:related} reviews related work and introduces DMR.
Section~\ref{sec:methodology} details the proposed methodology, while Section~\ref{sec:implementation} describes its integration into the DMR framework and the design of DMRv2.
Section~\ref{sec:evaluation} presents an experimental evaluation on three clusters ranked within the top 50 of the TOP500 list at the time of the experiments (October 2025).
Section~\ref{sec:discussion} discusses limitations and future work, and Section~\ref{sec:conclusions} concludes.

\section{Related Work and Background}\label{sec:related}
\begin{figure}[tbp]
    \centering
    \includegraphics[clip,width=0.75\linewidth,trim={0.2cm 0.2cm 0.2cm 0.2cm}]{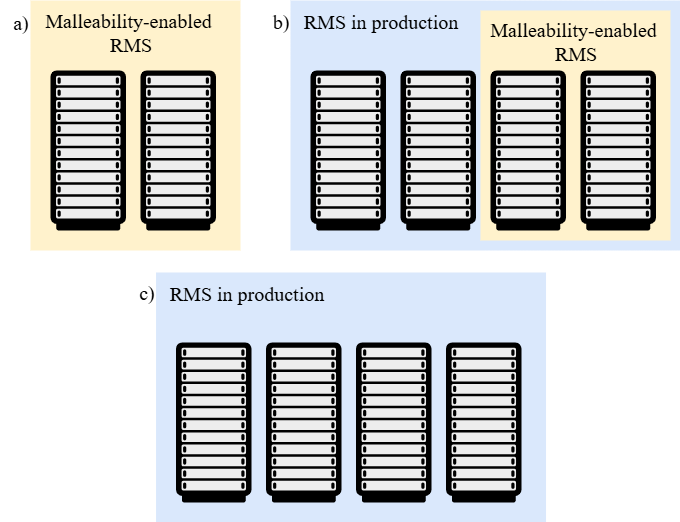}
    \caption{RMS deployments for dynamic resource management. a) and b) are common configurations in the literature, while c) is the target of this work.}
    \label{fig:cluster}
\end{figure}

Figure~\ref{fig:cluster} illustrates the three deployment scenarios for dynamic resource management in HPC clusters. In Figure~\ref{fig:cluster}a, the entire cluster runs a \emph{customized} RMS deployment (e.g., a modified Slurm), which in principle is ideal for enabling dynamic resource management but in practice is restricted to small-scale experimental testbeds. To the best of our knowledge, there is no documented case of a dedicated, malleability-enabled cluster operating at meaningful scale; deploying such a system as a production service would be operationally challenging due to the administrative, maintenance, and risk concerns discussed in Section~\ref{sec:intro}.

The configuration most commonly found in the dynamic resource management literature corresponds to Figure~\ref{fig:cluster}b~\cite{Lemarinier2016, Ribeiro2013, Sudarsan2007, ElMaghraoui2007, Gupta, Martin2013, compres_infrastructure_2016}. A production cluster runs a standard RMS instance, and dynamic features are enabled only within a subset of reserved nodes by launching a nested, malleability-enabled RMS as part of a job. When such a job starts, the nested RMS is configured for the assigned resources, and customized daemons are deployed accordingly (e.g., while a node runs the controller such as \texttt{slurmctld}, the remaining nodes run per-node daemons such as \texttt{slurmd}).

In this regard, Elastic MPI~\cite{compres_infrastructure_2016} was proposed as an infrastructure and API extension for malleable MPI applications, based on Slurm and the MPI implementation MPICH.\footnote{\url{http://www.mpich.org}.} In Elastic MPI, both Slurm and MPICH are extended with additional functionality to support job reconfiguration, enabling Slurm to control the creation and removal of MPI processes while managing resource allocations. Similarly, Chadha \emph{et al.}~\cite{chadha_extending_2021} extend Slurm to support malleable jobs via the Invasive MPI framework, following the same overall design philosophy. 
None of these approaches, however, leverage malleability in conjunction with the site-wide RMS on production supercomputers: they deploy a full instance of a customized RMS over a subset of nodes, either as a separate cluster or nested within an existing instance, which limits applicability where administrators require vanilla configurations.

Beyond RMS-level modifications, some efforts combine workflow management systems (WMSs) with RMSs to attain elasticity. Bhattarai \emph{et al.}~\cite{bhattarai_dynamic_2024}, for example, propose a fine-grained elastic resource reallocation mechanism that couples the Parsl WMS with a customized PMIx-enabled Slurm. Houzeaux \emph{et al.}~\cite{houzeaux_dynamic_2022} extend COMPSs to interact with a vanilla Slurm deployment on a dedicated 38-node cluster, enabling resource reallocation within a job.

The Dynamic Management of Resources (DMR) framework~\cite{iserte_agut_high-throughput_2018} enables MPI applications to adapt their resource usage at runtime by acting as a bridge between the message-passing runtime (MPI plus its process manager, e.g., MPICH + Hydra, Open MPI + PRRTE) and the resource management system (e.g., Slurm). At a high level, DMR coordinates process management and resource reconfiguration across these layers, allowing applications to change the number of processes or nodes without being tightly coupled to the underlying scheduler mechanisms. In existing deployments, DMR has been used primarily in controlled environments where a customized RMS instance (often based on Slurm) runs on a dedicated set of nodes (see Figure~\ref{fig:cluster}a--b). 
DMR relied on Slurm’s ability to resize running jobs to implement malleability, although recent Slurm versions have removed support for expanding a running jobs, further constraining malleable use cases~\cite{iserte_resource_2025}.

From the application perspective, DMR exposes an API that lets users express where and how reconfigurations may occur, typically by (i) initializing the DMR runtime alongside MPI, (ii) periodically checking whether a reconfiguration should be triggered according to a given policy, (iii) executing user-defined data redistribution routines when a reconfiguration happens, and (iv) finalizing the DMR runtime at the end of the execution. These operations allow applications to remain largely agnostic to how resources are obtained or released by the underlying system.

In contrast to prior approaches, this work targets, for the first time in the literature, the configuration depicted in Figure~\ref{fig:cluster}c. Our methodology does not modify the configuration of the site-wide RMS orchestrating the cluster. Instead, it exploits user-level allocation mechanisms in Slurm together with the PMIx Reference RunTime Environment (PRRTE)\footnote{\url{https://docs.prrte.org/}.} to dynamically manage allocations and MPI processes in a manner compatible with production supercomputers. As a result, malleable applications developed with the new DMR API can be executed dynamically on production clusters using unmodified, vanilla Slurm installations, without requiring nested RMS deployments or administrator intervention.

\section{Non-invasive Resource Management in MPI Malleability}\label{sec:methodology}
This section describes the proposed non-invasive methodology for enabling MPI malleability in production clusters.

Conventional wisdom in HPC assumes that dynamic resource management requires an RMS capable of resizing jobs dynamically. This work challenges that assumption: the proposed methodology removes the dependency on RMS-level dynamic resizing. Instead, it interacts directly with Slurm’s C-based API to request and release resources at the user level and orchestrates MPI processes across multiple independent Slurm jobs.
When using Open MPI\footnote{\url{https://www.open-mpi.org/}}, DMRv2 dynamically resizes the PRRTE-based distributed virtual machine (DVM) during reconfigurations.

For job expansions, the methodology follows a common strategy in the bibliography~\cite{aliaga_survey_2022}. 
It relies on \texttt{MPI\_Comm\_spawn}, which creates a new MPI communicator and connects it to its parent communicator, enabling communication among both sets of processes. 
Particularly, in Open MPI 5, PRRTE automatically expands the DVM to accommodate the new process layout. Since PRRTE currently cannot shrink the DVM once it has been deployed on a node, an alternative mechanism is provided: a wrapper script restarts the DMR-enabled application, including the DVM, under the new configuration. 
PRRTE can be configured to tolerate daemon failures, enabling the application to continue when unused nodes are removed through the Slurm API and parts of the DVM disconnect, as is typical in production systems. 
In such a configuration, PRRTE provides an alternative to the wrapper script when in-memory reconfiguration is desired. 
We anticipate that, under a PRRTE configuration that natively supports shrinking the DVM.%, DMR could seamlessly integrate with future PRRTE releases that offer this capability.

The expansion operation proceeds as follows:
\begin{enumerate}
    \item The user launches the DMR-enabled application with \texttt{mpirun} from a Slurm job (e.g., via \texttt{sbatch}). This job is regarded as the ``parent'' and runs for the duration of the DMR-enabled application.
    \item DMR determines the new resource layout according to the selected reconfiguration policy. %For example, if the execution reports communication efficiency above a predefined threshold, DMR decides to request additional resources. 
    \item DMR uses Slurm API calls to submit an ``expander'' job whose wallclock matches the remaining requested time of the parent job. This expander job is launched with a script that periodically checks that the parent job is still alive and requests the difference between the current and the desired node count.
    \item Once DMR detects that the necessary resources have been allocated, the application execution is suspended.
    \item The resources assigned to the expander job are recorded by DMR and the DVM is expanded accordingly using MPI spawn or checkpoint-restart mechanisms, in both cases using \textit{Secure SHell (SSH)} bootstrapping.
    \item Data redistribution is performed using application-specific logic provided by the user and the execution is resumed from the point at which it was halted.
\end{enumerate}

To shrink applications, the methodology uses a similar procedure, adapted to resource deallocation. DMR ensures that resources marked for release are properly relinquished before the application continues execution. Since not all Slurm configurations allow dynamic resource removal, DMR is designed to handle both of the following cases:
\begin{itemize}
    \item If Slurm supports job size reduction, DMR specifies directly the number of nodes to remove.
    \item Otherwise, shrinking is performed in whole-job units: DMR terminates only expander jobs created during execution; if no such jobs exist, shrinking is not possible.
\end{itemize}
This configurability enables DMR to maintain compatibility across a wide range of Slurm deployments, ensuring robust behavior regardless of the underlying scheduler capabilities.

The shrinking operation proceeds as follows:
\begin{enumerate}
    \item The application is launched inside a Slurm job, following the same procedure as in the expansion case.
    \item As in step~2 of the expansion operation, DMR determines the new resource layout according to the selected reconfiguration policy.
    %the execution is monitored; in this case, communication efficiency falling below a predefined threshold may trigger a decision to shrink.
    \item Using either MPI spawn or checkpoint-restart mechanisms, DMR instantiates a new set of processes with fewer ranks than the original communicator.
    \item Data redistribution is performed using application-specific logic provided by the library user.
    \item On the newly created processes, DMR identifies Slurm resources that are not associated with the new MPI communicator and invokes Slurm API calls to resize or terminate the corresponding expander jobs or, if necessary, to resize the parent job.
    \item The execution is resumed from where it was halted.
\end{enumerate}

Overall, the methodology is built around a \textit{respawning} 
mechanism, in which each reconfiguration restarts the process set to
ensure clean and deterministic transitions between configurations.
Alternative strategies, such as merging communicators or using
auxiliary threads to handle process spawning~\cite{martin-alvarez_dynamic_2023} or leveraging MPI Sessions~\cite{huber2024designprinciplesdynamicresource} %huber_bridging_2025} 
for dynamic process adaptation have already been explored in the literature.

\section{DMRv2 Design and Implementation}
\label{sec:implementation}
This section presents the DMRv2 API and its implementation following the non-invasive resource management methodology introduced before.

\begin{figure}[tbp]
    \centering
    \includegraphics[clip,width=0.7\linewidth,trim={6.3cm 12.2cm 5.8cm 4.8cm}]{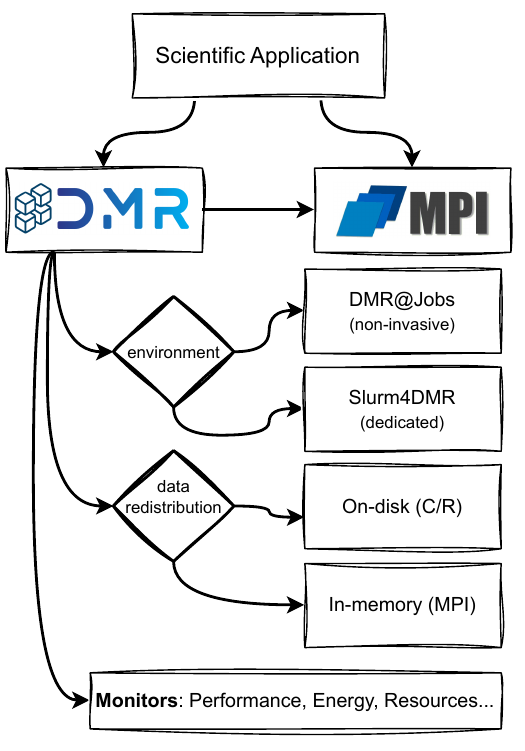}
    \caption{DMRv2 design and working modes.}
    \label{fig:dmr}
\end{figure}

The DMRv2 API lets users express malleable behavior with a syntax that is agnostic to the underlying reconfiguration mechanism, supporting both production deployments (DMR@Jobs) and controlled testbeds (Slurm4DMR). Figure~\ref{fig:dmr} illustrates the refactored design: scientific applications use MPI and DMR together, with MPI handling communication and process management, and DMR orchestrating job reconfigurations. In recent Open MPI releases (version~5), process management is delegated to the runtime PRRTE, which reacts to application events (such as launches) or explicit MPI calls; DMRv2 builds on this process-management capability.

The revised DMR API enables users to specify the data redistribution operations to be executed during reconfiguration. These user-provided functions may implement checkpoint-restart or in-memory communication to map application data onto the new process layout. The result is a DMRv2 API through which users can configure the mechanism underlying dynamic execution (e.g., Slurm4DMR vs. DMR@Jobs) without modifying their application’s malleable code. For example, Slurm4DMR can be used during development, to test malleable applications in a controlled environment where resources are pre-allocated and can be acquired or released on demand, while the same application code may later be deployed in production using DMR@Jobs.

In legacy DMR~\cite{iserte_dmr_2018}, jobs could only be submitted to the custom resource manager environment (Slurm4DMR). As a result, once Slurm4DMR had been launched under a fixed reservation, resource requests could be satisfied almost instantly. In production supercomputers, however, resources are heavily contended; job submissions may wait in queues for non-trivial and non-deterministic periods of time, and a global view of cluster status is often unavailable to users (e.g., on the MareNostrum 5 (MN5) supercomputer that we use in our experiments), depending on the Slurm configuration. Consequently, in the DMR@Jobs approach, it is not reasonable to block application execution by default while waiting for resources. Instead, DMR@Jobs hides most of the resource acquisition latency by allowing the application to continue executing while resource requests for expansions are pending.

To support this new usage model while maintaining a simple user experience, the DMR API has been reworked so that reconfiguration is primarily driven by the following functions:
\begin{itemize}
    \item \texttt{\textbf{dmr\_init(argc, argv)}}: Initializes DMR internal data structures and determines whether the current process is part of a restarted configuration.
    \item \texttt{\textbf{dmr\_check(DMRSuggestion)}}: Requests a resource reallocation according to the hints or policies encoded in the \texttt{DMRSuggestion} argument, or checks the status of a previously issued request.
    \item \texttt{\textbf{dmr\_reconfigure()}}: Triggers a scheduled reconfiguration coordinated by \texttt{dmr\_check(...)}, or finalizes an ongoing reconfiguration after initialization. In many cases, users do not need to call this function explicitly.
    \item \texttt{\textbf{dmr\_finalize()}}: Releases internal DMR data structures and performs final cleanup.
\end{itemize}

Each function returns a value of type \texttt{DMRAction}, indicating any required follow-up action. For example, if additional resources have been acquired and the user calls \texttt{dmr\_check(...)}, the return value may be \texttt{DMR\_RECONF}, suggesting that the application should, at a convenient synchronization point, invoke \texttt{dmr\_reconfigure()}. To streamline common usage patterns, DMRv2 provides a helper macro \texttt{DMR\_AUTO}, which accepts a \texttt{DMRAction} (or a DMR function) and automatically executes the required follow-up actions, including application-specific data redistribution functions when supplied. Its syntax is \lstinline| DMR_AUTO(  dmr_func,  redist_func,  restart_func,  finalize_func  )| where:

    \begin{itemize}
        \item \texttt{\textbf{dmr\_func}}: Any of the DMR functions above or their returned \texttt{DMRAction}.
        \item \texttt{\textbf{redist\_func}}: User-provided function for data redistribution or checkpointing before reconfiguration.
        \item \texttt{\textbf{restart\_func}}: User-provided function to restore data after reconfiguration, if needed.
        \item \texttt{\textbf{finalize\_func}}: User-provided function to release application data after redistribution or at finalization.
    \end{itemize}

\begin{figure}[tbp]
\begin{lstlisting}[language=C, caption=Skeleton of an example malleable code using the DMRv2 API., label=lst:dmrpseudo, captionpos=b, basicstyle=\fontsize{7}{9}\selectfont\ttfamily]
int main(int argc, char *argv[]) {
    MPI_Init(...);
    DMR_AUTO(dmr_init(argc, argv), (void)NULL, data_receive(...), (void)NULL);
    while(time_steps) {  
        DMR_AUTO(dmr_check(...), data_send(...), (void)NULL, data_cleanup(...));
        compute(); // Application-specific computation
    }
    DMR_AUTO(dmr_finalize(), (void)NULL, (void)NULL, data_cleanup(...));
    MPI_Finalize();
    return EXIT_SUCCESS;
}
\end{lstlisting}
\end{figure}

The macro expands to a \texttt{switch} statement that calls the appropriate follow-up actions based on the \texttt{DMRAction} value. Unused handlers may be provided as \texttt{(void)NULL}, in which case these are ignored. Listing~\ref{lst:dmrpseudo} shows the structure of a DMRv2-enabled malleable application. Line~2 initializes the MPI environment. Line~3 initializes the DMR environment and, when applicable, loads data on derivative processes via \texttt{data\_receive(...)}, while the send and finalize handlers are omitted at this stage. The \texttt{data\_receive(...)} function is user-defined and may use MPI to receive data from original processes or I/O to restore state from checkpoints.

The example corresponds to an iterative application in which the new API is invoked at the start of each time step. On line~6, \texttt{dmr\_check(...)} determines whether reconfiguration is appropriate based on user hints and, for expansions, whether additional resources have been acquired. If a reconfiguration is triggered, \texttt{data\_send(...)} performs data redistribution from the original processes, after which those processes terminate. An optional function \texttt{data\_cleanup(...)} is then called to clear any application-specific state on those processes. Once the overall execution completes, \texttt{dmr\_finalize()} on line~8 releases DMR resources, followed by a final call to \texttt{data\_cleanup(...)}.

In legacy DMR, reconfiguration decisions were tightly coupled to the custom resource manager (Slurm4DMR), and changing the reconfiguration strategy required recompilation. With DMR@Jobs, there is no requirement for a custom resource manager, so a new mechanism was needed to express and combine reconfiguration strategies at runtime. To this end, DMRv2 introduces the enum \texttt{DMRSuggestion}, which is passed to \texttt{dmr\_check(...)}, allowing users to suggest actions to DMR. In the simplest case, suggestions directly indicate how to reconfigure:
\begin{enumerate}
    \item \texttt{SHOULD\_SHRINK}: Shrink the job and continue execution with fewer resources.
    \item \texttt{SHOULD\_EXPAND}: Expand the job and continue execution with additional resources.
    \item \texttt{SHOULD\_STAY}: Keep the current configuration.
\end{enumerate}

The target number of nodes or processes can be fixed at compile time or adjusted dynamically at runtime, depending on application needs. More sophisticated behaviors are expressed as policies, which internally translate to the basic \texttt{SHOULD\_EXPAND}, \texttt{SHOULD\_SHRINK}, or \texttt{SHOULD\_STAY} suggestions. DMRv2 supports policies inherited from legacy DMR (which require Slurm4DMR) and new policies that may operate without privileged access to cluster-wide RMS information, which may not be available in all Slurm configurations.
\begin{enumerate}
    \item \texttt{ROUND\_POLICY}: Cycles between a minimum and maximum resource count by repeatedly growing (multiplying resources) up to a specified maximum and then shrinking to a specified minimum; this is particularly useful during development and testing phases.
    \item \texttt{CE\_POLICY}: Adjusts the allocation to track a target communication efficiency\footnote{\url{https://pop-coe.eu/node/69}} (CE), using the TALP performance monitor~\cite{lopez_talp_2021} to measure the fraction of time spent in communication versus computation; the number of ranks is adapted linearly as a function of the deviation from the target CE, with larger deviations triggering more aggressive expansions or shrinkages The integration between TALP and DMR is described in detail in~\cite{iserte_mpi_2025}.
    \item \texttt{QUEUE\_POLICY}: Adapts the allocation based on the job queue, using RMS information about queued jobs and idle nodes to improve cluster productivity in terms of completed jobs per unit time. This policy is only usable when Slurm information is exposed to the user, and currently it requires Slurm4DMR.
\end{enumerate}

An advantage of the \texttt{DMRSuggestion} abstraction is that reconfiguration strategies may be combined at runtime without recompilation. For example, an application may use \texttt{CE\_POLICY} during most of the simulation to maintain a target CE and then switch to \texttt{SHOULD\_SHRINK} near the end to reduce resources for post-processing.

A large share of production scientific HPC applications are iterative simulations or solvers that repeatedly execute similar computation–communication patterns over many iterations or timesteps~\cite{totoni, xie, ekelund}. For this reason, we provide the helper macro to simplify the most common case. More advanced usage is also supported: users can directly leverage the DMR API to implement fine-grained, layered scalability (for example, allowing an individual solver component to dynamically grow or shrink resources in response to data-driven needs).

Together, these design and implementation choices allow DMRv2 to support both the Slurm4DMR and DMR@Jobs deployment models with a single, unified API, enhancing portability and easing adoption in production environments.

\section{Performance Evaluation}\label{sec:evaluation}
We evaluate the presented methodology implemented in DMRv2 with experiments designed to positively answer the following questions:
\begin{enumerate}[label=\alph*)]
  \item Can DMRv2 support production-grade malleable applications using both \textbf{checkpoint/restart and in-memory data redistribution}?
  \item Does the approach enable the same malleable code to run in both controlled environments and production systems bridging the gap between research \textbf{prototypes and operational HPC platforms}?
  \item Does the methodology operate robustly \textbf{across supercomputers and heterogeneous} CPU–GPU resources?
  \item Is the new API practical to integrate into large HPC codes, providing potential for \textbf{broad adoption by the community}?
\end{enumerate}

The core software stack is kept consistent across platforms: Open MPI~5.1.0a1, PRRTE~5.0.0a1, OpenPMIX~7.0.0a1, and DLB~3.5.0.

\subsection{Experimental setup and configuration}
We use two representative scientific applications. The first is a malleable version of Alya~\cite{iserte_malleable_2025}, a high-performance computational mechanics code for multiphysics problems~\cite{Vazquez15d}. The dataset models a turbulent flow over a sphere with 128 million elements solved with a fractional-step method for the incompressible Navier–Stokes equations and a conjugate gradient solver for the pressure correction equation~\cite{LEHMKUHL201951}. Alya includes a robust C/R mechanism, which we exploit to perform job reconfigurations during execution; the C/R strategy is designed to remain parallel and independent of the MPI process count throughout the workflow, from preprocessing (MPI-IO restart reading and parallel partitioning) to checkpointing (MPI-IO restart writing)~\cite{houzeaux_dynamic_2022}. In all Alya experiments, DMR is configured with \texttt{CE\_POLICY}, targeting a communication efficiency of 70\%.

The second application is a CUDA-enabled version of the MPDATA kernel~\cite{rojek_parallelization_2015}, a core component of the EULAG multiscale fluid solver responsible for advecting a non-diffusive quantity in a flow field through iterative time-stepping. A malleable variant based on OmpSs was previously proposed to improve energy efficiency~\cite{iserte_study_2020}; here, we port this application to DMRv2. The step-based structure and near-linear scalability of MPDATA make it representative of a broad class of stencil-like HPC workloads, and it has been widely used in prior malleability studies~\cite{huber2025dynamic, iserte_mpi_2025}. We use the ``ball'' dataset with a grid of $4096\times1024\times64$ cells; MPDATA employs in-memory data redistribution via MPI communication during job reconfigurations. In all MPDATA experiments, DMR uses \texttt{ROUND\_POLICY} over a node range of 2--16.

Table~\ref{tab:exps} summarizes the experimental configurations. Together, they exercise both data redistribution mechanisms (C/R and in-memory), both deployment modes (Slurm4DMR and DMR@Jobs), and both CPU-only and heterogeneous CPU–GPU platforms. Alya runs on CPU-only partitions to showcase checkpoint/restart-based malleability, while MPDATA runs on GPU-accelerated systems to validate in-memory redistribution under heterogeneous hardware.

\begin{table}[tbp]
    \centering
    \caption{Overview of experiments by supercomputer, environment, and application.}
    \scalebox{0.71}{%
    \begin{tabular}{l c c || c}
        \toprule
        Supercomputer (HW)\,\textbackslash\,Env. & Slurm4DMR (Controlled) & DMR@Jobs (Production) & Location \\ \midrule
        Leonardo (CPU)      & --                 & Alya (C/R)          & Section~\ref{subsec:leo}      \\
        MN5 ACC (CPU--GPU) & MPDATA (in-memory) & MPDATA (in-memory)  & Section~\ref{subsec:mn5acc}   \\
        MN5 GPP (CPU)      & Alya (C/R)         & Alya (C/R)          & Section~\ref{subsec:mn5gpp}   \\
        \bottomrule
    \end{tabular}
    }
    \label{tab:exps}
\end{table}

Because each experiment performed is highly resource intensive---consuming thousands of CPU hours with non-negligible monetary and environmental cost---it is important to carefully select the configurations to execute. At the scales considered here, execution times are sufficiently stable that single runs provide representative results, in contrast to microbenchmarks (e.g., bandwidth or latency) where variability may strongly affect outcomes. Consistent with common practice in the state of the art on malleability~\cite{aliaga_survey_2022, tarraf_malleability_2024}, our evaluation therefore uses a single execution per configuration.
Furthermore, we run on shared production systems, queue waiting times, node placement, and interference from co-running jobs vary between executions. As a result, repeated runs of the same configuration may exhibit different detailed timelines, even though overall trends remain comparable.
Nevertheless, this is not affecting the narrative of this work since \textbf{we are not re‑evaluating the benefits of MPI malleability} itself---which have been studied extensively (see Section~\ref{sec:related})---but on, for the first time, \textbf{assessing a deployment methodology} that makes existing malleability techniques usable \textbf{on production HPC platforms} without scheduler modifications.

\subsection{Leonardo -- TOP500 \#10}\label{subsec:leo}
This experiment evaluates how DMRv2 drives Alya towards efficient node allocations on a production CPU-only system, starting from contrasting initial sizes.

The pre-exascale EuroHPC Tier-0 supercomputer Leonardo\footnote{\url{https://www.hpc.cineca.it/systems/hardware/leonardo}} comprises two custom-designed compute blade types, exposed as separate partitions. In this work, the DMR-enabled version of Alya is deployed on the ``Data Centric General Purpose (DCGP)'' partition, which consists of 1,536 nodes, each equipped with two Intel Sapphire Rapids processors (112 cores total) and 512~GB of DDR5 memory, interconnected via a 200~Gbps NVIDIA Mellanox HDR InfiniBand network and managed by Slurm~23.11.10. Due to time allocation constraints on Leonardo, experiments on this system focus exclusively on assessing the proposed production-RMS methodology; comparisons against custom RMS-based approaches are deferred to the remaining testbeds.

\begin{figure*}
    \centering
    \includegraphics[width=0.9\linewidth]{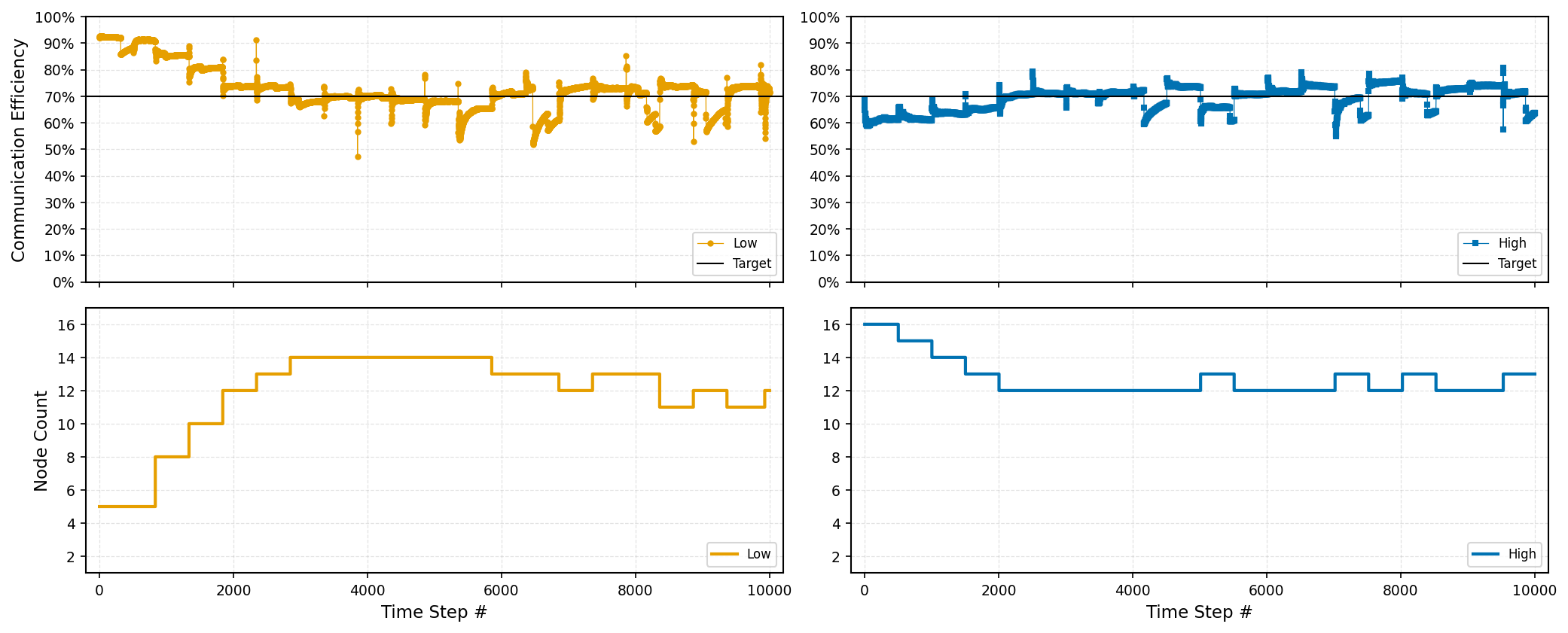}
    \caption{CE (top) and node count (bottom) per time step of Alya malleable executions on Leonardo. Left plots correspond to the \textit{low} job and right plots
    to the \textit{high} job. In the top graphs, the target CE is shown with a black horizontal line at 70\%.}
    \label{fig:leonardo-alya}
\end{figure*}

Figure~\ref{fig:leonardo-alya} shows the execution of two Alya instances on the DCGP partition. Both runs use the same problem setup and DMR configuration, differing only in their initial node allocation: one starts with 5 nodes (\textit{low}) and the other with 16 nodes (\textit{high}), as shown in the bottom plots. The top plots display the evolution of CE at each time step: the \textit{low} job initially exhibits a high CE, indicating low communication overhead relative to computation, whereas the \textit{high} job starts with low CE due to excessive communication caused by the larger initial node count. These communications appear mainly in the iterative solver phase of the pressure correction equation, here the conjugate gradient.

As the simulation progresses, both jobs converge towards similar configurations. The \textit{high} job stabilizes around time step~2{,}000, while the \textit{low} job reaches a steady regime around time step~3{,}000, with CE curves displaying periodic spikes that correspond to the start of each new measurement interval. DMR employs an \textbf{inhibition mechanism that prevents reconfigurations for a fixed number of time steps}---500 in these experiments---regardless of whether a reconfiguration occurs; CE is evaluated at the end of each inhibition period using the average over that interval, which makes early samples in each period more sensitive to outliers and increasingly stable as more measurements are accumulated.

Once stabilized, the \textit{high} job remains at 12--13 nodes, while the \textit{low} job fluctuates between 11–14 nodes. These small oscillations arise from two main factors: (1) variability in the number of inner iterations executed by the solver at each time step, and (2) non-deterministic congestion in the shared cluster network, both of which naturally perturb CE and thus the resizing decisions. The \texttt{CE\_POLICY} may be tuned via a tolerance parameter that controls how aggressively resources are added or removed in response to deviations from the target CE; for instance, previous work on malleable Alya used an 80\% CE target and a 2{,}500 time-step inhibition period for runs exceeding 90{,}000 time steps~\cite{iserte_malleable_2025}. In this study, such policy-level refinements are considered out of scope; the objective is to demonstrate that DMR can dynamically steer jobs toward more efficient configurations on a production system like Leonardo, regardless of the initial allocation or intermediate behavioral variability.

\subsection{Marenostrum 5 ACC -- TOP500 \#14}\label{subsec:mn5acc}
This experiment evaluates how the DMRv2-enabled MPDATA-GPU application, which uses in-memory data redistribution, can be executed seamlessly on a heterogeneous CPU–GPU cluster in both controlled and production environments, while comparing node-hour utilization in the two scenarios.

MareNostrum 5 (MN5) is a pre-exascale system integrated into the EuroHPC-JU infrastructure\footnote{\url{https://bsc.es/marenostrum/marenostrum-5}}. Its accelerated partition (ACC)\footnote{\url{https://bit.ly/mn5-acc}} comprises 1,120 Intel Xeon Sapphire Rapids-based nodes, each with two Intel Xeon Platinum 8460Y+ processors (80 cores in total), 512~GB of DDR5 memory, and four NVIDIA H100 GPUs, interconnected via four ConnectX-7 NDR200 InfiniBand NICs (up to 800~Gbit/s); MN5 ACC uses Slurm~23.02.7 as its RMS, and the experiments rely on CUDA~12.8 with a CUDA-aware Open MPI build.

For this platform, a DMRv2-capable version of the malleable GPU-enabled MPDATA application~\cite{iserte_study_2020} is developed. During initial development and debugging, DMR is configured with \texttt{ROUND\_POLICY} on top of Slurm4DMR, iterating over a simple sequence of node allocations (e.g., $1, 2, 4, 1, 2, \dots$), which is convenient for rapidly exercising expansion and shrink operations. Once the application behaves as expected in this controlled environment, it is deployed on the production MN5 ACC system under DMR@Jobs, configuring \texttt{ROUND\_POLICY} with a node range of 
$[2{-}16]$.

\begin{figure}
    \centering
\includegraphics[width=0.9\linewidth]{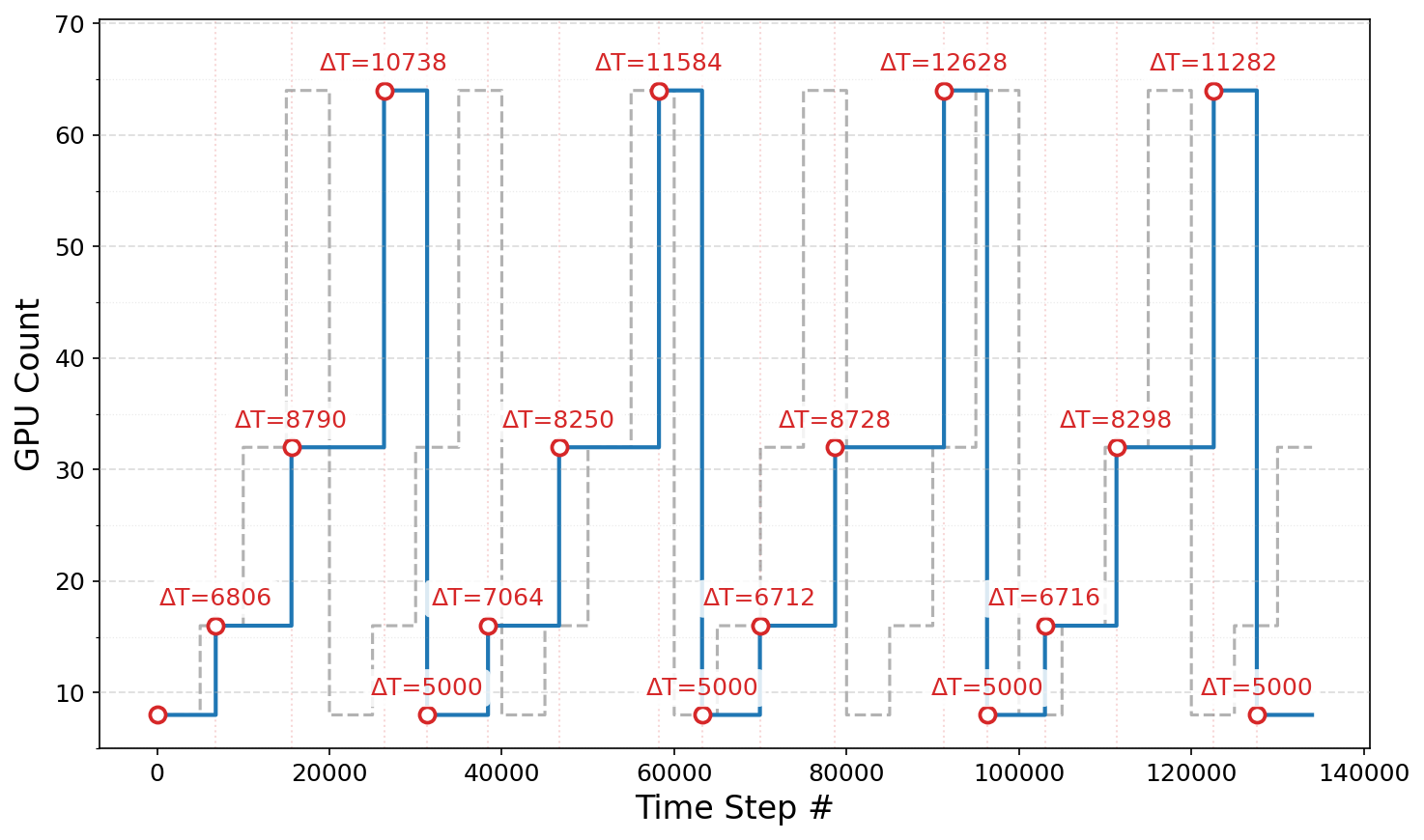}
    \caption{MPDATA malleable on MN5 ACC with \texttt{ROUND\_POLICY} indicating the number of time steps needed for each reconfiguration until resources are available. The gray dashed line shows a Slurm4DMR-based execution and the solid blue line shows a DMR@Jobs-based execution. }
    \label{fig:mn5-acc}
\end{figure}

Figure~\ref{fig:mn5-acc} reports the evolution of the number of allocated GPUs (four per node) over time, showing the number of time steps among successive reconfigurations. An inhibition period of 5{,}000 time steps is used, establishing the minimum spacing among reconfigurations. In the Slurm4DMR case (gray dashed line), which runs without external contention, reconfigurations occur exactly every 5{,}000 time steps, as configured. In contrast, the DMR@Jobs execution on the production system (solid blue line) exhibits variability in the number of time steps required for expansions, reflecting asynchronous waits for additional resources under the control of the system-wide Slurm scheduler, whereas shrink operations can release resources immediately and thus respect the fixed 5{,}000-step interval.

Under heavier system load, expansions would be expected to require even more time steps, as resource acquisition remains asynchronous, but resource release would still occur in effectively constant time, independent of congestion. From a cost perspective, running in a dedicated Slurm4DMR environment requires continuously reserving the maximum number of nodes plus an additional controller node; in the MPDATA case, this configuration uses 17 nodes for 40 minutes, corresponding to 11.5 node-hours. The DMR@Jobs run on MN5 ACC, by contrast, completes in 41 minutes while only allocating the resources actually needed at each stage, for a total of 3.0 node-hours, i.e., a 74\% reduction in node-hour consumption. 
This provides a strong incentive for HPC centers to adopt MPI malleability to improve resource utilization and, to the best of our knowledge, constitutes the first practical demonstration that MPI malleability can be effectively exploited on top of malleability-oblivious RMSs in a production GPU-accelerated system.

\subsection{Marenostrum 5 GPP -- TOP500 \#45}\label{subsec:mn5gpp}
Similarly to the previous experiment in Section~\ref{subsec:leo}, this experiment evaluates how DMRv2’s CE-based resizing behaves for a checkpoint/restart application in a CPU-only setting, focusing on the transition from a controlled to a production environment and its impact on node-hour usage for under- and over-provisioned Alya runs.

\begin{figure*}[tbp]
    \centering \includegraphics[width=0.9\linewidth]{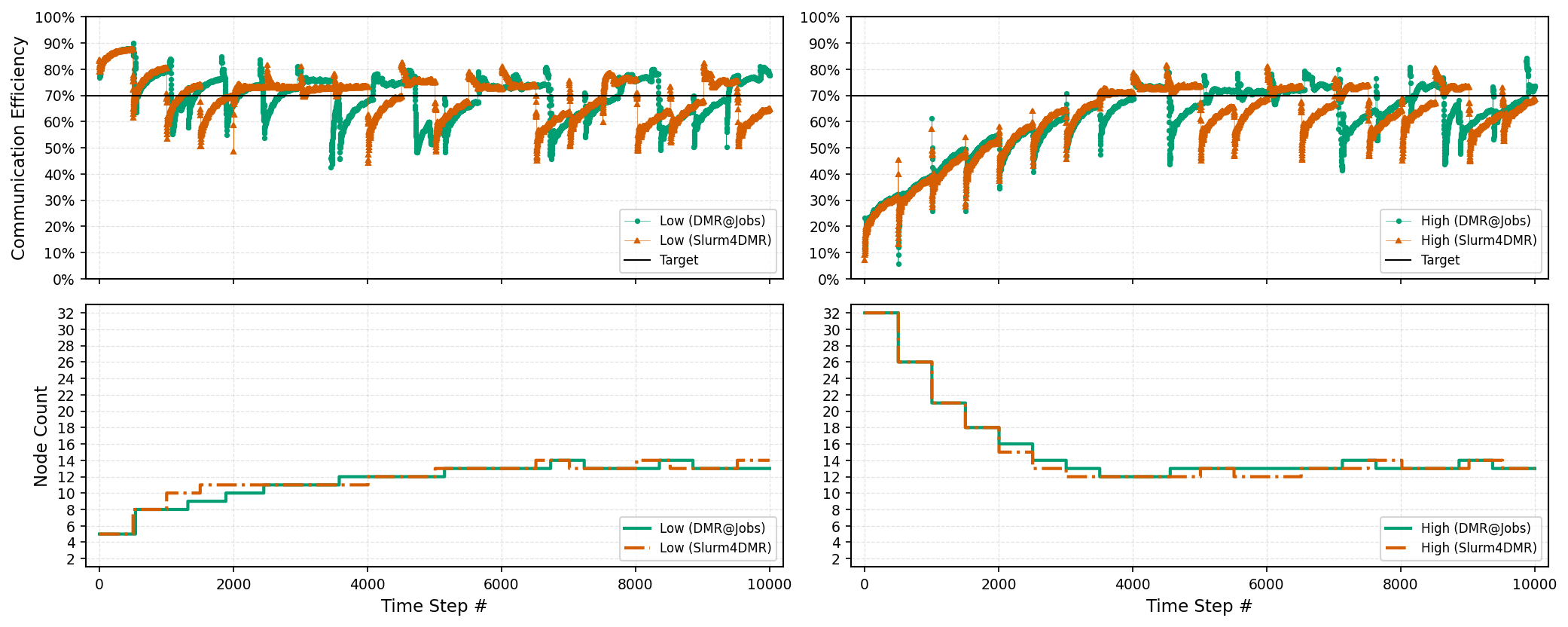}
    %\caption{DMRv2-enabled malleable Alya executions on MN5, GPP.}
\caption{CE (top) and node count (bottom) per time step of Alya malleable executions on MN5 GPP. Left plots correspond to the \textit{low} jobs and right plots represent the \textit{high} jobs. In the top graphs, the target CE is shown with a black horizontal line at 70\%.}
    \label{fig:mn5-gpp}
\end{figure*}

The MareNostrum 5 general-purpose partition (GPP)\footnote{\url{https://bit.ly/mn5-gpp}} comprises 6,192 Intel Sapphire Rapids-based nodes, each equipped with two Intel Xeon Platinum 8480+ processors (56 cores at 2~GHz, 112 cores total) and 256~GB of DDR5 memory, interconnected by a 100~Gbit/s ConnectX-7 NDR200 InfiniBand network and managed by Slurm~23.02.7. Figure~\ref{fig:mn5-gpp} reports the behavior of \textit{low} and \textit{high} Alya jobs in terms of CE and allocated nodes per time step, mirroring the analysis in Section~\ref{subsec:leo} but now comparing controlled (Slurm4DMR) and production (DMR@Jobs) executions on the same system.

In these experiments, both environments use \texttt{CE\_POLICY} with a 70\% CE target. The \textit{low} jobs (left plots) start with 5 nodes, whereas the \textit{high} jobs (right plots) start with 32 nodes. Runs using DMR@Jobs and the system-level MN5 GPP Slurm are shown in green, while runs using the controlled Slurm4DMR environment appear in red. For the \textit{low} job, the red dashed line showing the node count under Slurm4DMR increases slightly faster than the green line for DMR@Jobs, which is expected, since Slurm4DMR experiences no external contention and can reconfigure instantly, while production expansions must compete for resources with other jobs. For the \textit{high} job, both approaches behave nearly identically up to about iteration 2,000, illustrating that shrinking decisions are consistent across environments; by around iteration 5,000, both converge to a stable configuration of roughly 13 nodes, regardless of the initial allocation or whether the job runs under Slurm4DMR or DMR@Jobs.

Table~\ref{tab:alya-comparison} quantifies the cost in terms of execution time and node-hours. In the controlled Slurm4DMR environment, an additional node is required for management, and hence the \textit{low} job uses 14+1 nodes for 2.68 hours (40.20 node-hours), while the production DMR@Jobs run completes in 2.80 hours with a varying allocation between 5 and 14 nodes for a total of 30.09 node-hours, a 25.10\% reduction. For the \textit{high} configuration, the controlled run uses 32+1 nodes for 2.48 hours (81.84 node-hours), whereas the production run finishes in 2.36 hours with 12--32 nodes (36.87 node-hours), a 55.15\% reduction. In practice, the exact target node count is rarely known \textit{a priori}, and additional nodes are typically reserved in controlled configurations to avoid resource shortages; the results show that DMR@Jobs can instead adjust allocations dynamically in a production environment, significantly lowering node-hour costs while converging to similar efficient configurations as the controlled Slurm4DMR deployment.

\begin{table}[tbp]
\centering
%\fontsize{7pt}{9pt}\selectfont
\caption{Cost comparison between controlled and production executions.}
\label{tab:alya-comparison}
\scalebox{0.72}{%
\begin{tabular}{@{}lccc ccc c@{}}
\toprule
 & \multicolumn{3}{c}{\textbf{Controlled (Slurm4DMR)}} 
 & \multicolumn{3}{c}{\textbf{Production (DMR@Jobs)}} 
 & \textbf{Cost Reduction in} \\
\cmidrule(lr){2-4} \cmidrule(lr){5-7}
 Job    & \textbf{Time (h)} & \textbf{\#Nodes} & \textbf{Cost (n-h)} 
        & \textbf{Time (h)} & \textbf{\#Nodes} & \textbf{Cost (n-h)} 
 & \textbf{Production (\%)} \\
\midrule
\textit{low}  & 2.68 & 14+1 & 40.20 & 2.80 & [5--14] & 30.09 & 25.10\% \\
\textit{high} & 2.48 & 32+1 & 81.84 & 2.36 & [12--32] & 36.87 & 55.15\% \\
\bottomrule
\end{tabular}
}
\end{table}

\subsection{Workload-level Evaluation}\label{subsec:workloads}
This experiment evaluates the behavior and overheads of DMRv2-enabled malleable jobs at the workload level, where many applications compete for resources on a shared production cluster.

While the previous experiments focus on single malleable jobs, in this subsection we assess the impact of DMRv2 at the workload level, where multiple jobs compete for resources on a production cluster. Unlike prior work in which workloads are mainly used to quantify the benefits of dynamic resource management against traditional rigid jobs~\cite{iserte_agut_high-throughput_2018,tarraf_malleability_2024}, here we analyze the new behaviors and features that malleable jobs exhibit when deployed directly in production environments. The goal is to understand how malleable jobs behave under non-controlled conditions, recognizing that each workload execution will differ because the cluster status is never identical among different runs.

For this purpose, a 50-job workload is generated in which each job instantiates the Alya case described previously in this section. In this study, the total number of time steps per job is limited to 800, and the inhibition period among potential reconfigurations in terms of time steps is selected uniformly random in the range [10–100]. Jobs are submitted with a node range from 2 to 32 (\texttt{-N2-32}), and their interarrival times are sampled uniformly in the interval [0–100] seconds; the target CE is set to 75\%.

\begin{figure}[tpb]
    \centering
    \includegraphics[width=0.95\linewidth]{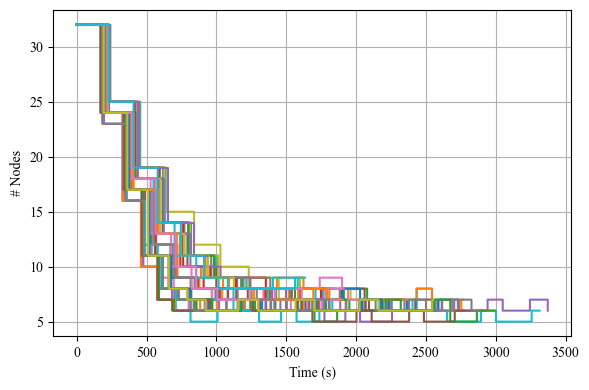}
    \caption{Node allocation over time for a 50-job Alya workload on MN5 GPP using DMRv2 and \texttt{CE\_POLICY} 75\%.}
    \label{fig:multijob10}
\end{figure}

Figure~\ref{fig:multijob10} shows the evolution of the node allocation for each job over its makespan (each line corresponds to one of the 50 jobs). 
In this experiment, all jobs are effectively started with the upper limit number of nodes, as determined by the MN5 GPP Slurm scheduler. Because all jobs simulate the same problem and use the same policy, these exhibit qualitatively similar resizing patterns; however, their trajectories differ due to variations in arrival times, inhibition periods, and the cluster state at the moment each reconfiguration is requested.

\begin{figure}[tpb]
    \centering
    \includegraphics[width=0.95\linewidth]{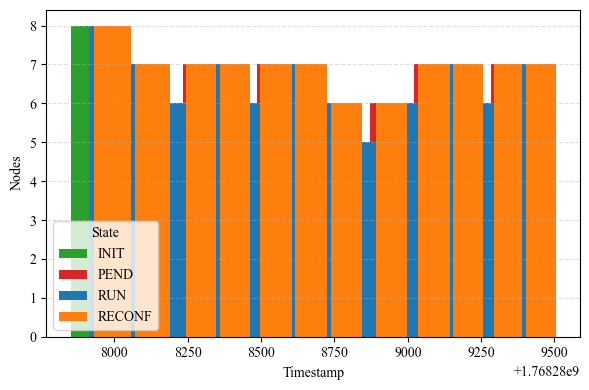}
    \caption{Example of the timeline of a job's stages.}
    \label{fig:pend-reconf}
\end{figure}

Notice that, because of the short inhibition periods, reconfigurations are triggered frequently throughout the workload. More realistic configurations, such as inhibition periods of 5,000 time steps, can be found in prior work~\cite{iserte_malleable_2025}. Although this configuration is intentionally unrealistic, it is insightful because it exposes many of the transitions among job states and generates more overhead statistics. For instance, Figure~\ref{fig:pend-reconf} depicts the detailed behavior of one of the jobs in the workload. 
Four states are distinguished to illustrate timing behavior: \textit{INIT} (initialization), \textit{PEND} (waiting time before resources are granted), \textit{RUN} (actual time step execution), and \textit{RECONF} (time spent resizing the job and reshaping the MPI process layout).

The experiment highlights that, with such short inhibition periods, reconfiguration time dominates execution, averaging 107.14 seconds in the \textit{RECONF} state. Moreover, the plot demonstrates that Alya continues executing while additional resources are pending during expansions---evident from overlapping \textit{RUN} and \textit{PEND} intervals---thus avoiding idle time while waiting for resources under the inherently uncertain conditions of a production cluster.

\section{Discussion}\label{sec:discussion}

This work presents the design and evaluation of an extensive dynamic resource management framework that allows developers to prototype dynamic applications in controlled environments and then deploy them on large-scale HPC systems with the very same malleable code. DMRv2 bridges parallel runtimes (e.g., PRRTE) and resource managers (e.g., Slurm) through a unified high-level API. Rather than assuming the presence of a malleability-aware RMS, the proposed methodology operates on top of unmodified production schedulers, using user-level interactions and multiple jobs to orchestrate expansions and shrinks. The evaluation on world-class supercomputers shows that this approach can dynamically steer scientific applications toward efficient configurations, achieving substantial reductions in node-hour consumption while preserving performance comparable to controlled custom-RMS configurations. 

Drawing on prior work in this area, we have repeatedly observed that, although in-memory data redistribution is typically more suitable for efficient job reconfiguration, developers are often reluctant to adopt such techniques when robust C/R mechanisms are already integrated into their applications. For this reason, DMRv2 deliberately supports both in-memory and C/R-based data redistribution, and exposes them through the same programming model; this dual support is essential to foster widespread adoption of malleability across diverse HPC codes and development cultures. In this sense, the contribution of DMRv2 is not to introduce yet another dynamic resource management API, but to provide an MPI malleability-aware programming model that is attentive to developers’ practical needs around data movement and that offers the  tools required to move from debugging malleable codes in controlled configurations to running these unchanged on production supercomputers. 

Despite dynamic resource management being widely studied~\cite{aliaga_survey_2022, tarraf_malleability_2024}, DMRv2 makes it more practical and accessible by offering a straightforward, end--to--end approach. At the system level, it can: (i) increase resource utilization by reducing fragmentation and right-sizing allocations over time, (ii) improve cluster productivity in terms of completed jobs per unit of time, (iii) reduce job waiting times by allowing jobs to start with fewer resources and expand later, and (iv) lower computational waste when guided by efficiency metrics such as communication efficiency. 
At the application level, DMRv2 is presented as a holistic tool that covers the full malleability operative, from development and testing in Slurm4DMR to deployment in production via DMR@Jobs, with an accessible API co-designed with domain scientists to match existing code structures. Beyond its direct use, this work has also influenced the surrounding software ecosystem: the groundwork laid by this work is currently leveraged in a community effort to extend PRRTE to become fully elastic under both malleability-oblivious and malleability-aware RMSs.

\section{Conclusions}\label{sec:conclusions}

This work demonstrates that dynamic resource management can be practically and competitively deployed on top of malleability-oblivious RMSs in production HPC environments, without requiring scheduler forks or intrusive configuration changes. The results dispel concerns that waiting for additional jobs during expansions would inherently negate the benefits of malleability, showing instead that dynamic resizing can reduce node-hour usage by substantial margins while converging to similar configurations as those in custom RMS environments. The experiments on Leonardo and MN5 confirm that the proposed methodology is robust across infrastructures and heterogeneous CPU-GPU platforms. 

The DMRv2 framework encapsulates these ideas in a single, portable API that supports both Slurm4DMR and DMR@Jobs modes, enabling developers to use the same malleable code in controlled and production settings and thereby lowering the barrier for adoption of malleability techniques in production HPC deployments. Although the current implementation targets Slurm and Open MPI/PRRTE, the methodology is not tied to these specific systems; ongoing work includes integration with other RMS and MPI implementations such as Flux~\cite{flux}, OAR~\cite{10.1007/978-3-031-90203-1_34}, and MPICH\cite{mpich2025}, as well as further extensions toward CI/CD integration and HPC-QC convergence scenarios. Overall, the proposed approach contributes a practical path for seamlessly bringing scientific applications and dynamic resource management from experimental prototypes into production supercomputers.

\section*{Acknowledgment}
This work/publication is promoted by the Barcelona Zettascale Laboratory, backed by the Ministry for Digital Transformation and of Public Services, within the framework of the Recovery, Transformation, and Resilience Plan - funded by the European Union - NextGenerationEU.

Language polishing was performed using an AI language model, with subsequent thorough review and editing by the authors. The authors take full responsibility for the final manuscript content.

\bibliographystyle{IEEEtran}
\bibliography{bib/bib}

\end{document}